\documentclass{svproc}
\usepackage{url}
\PassOptionsToPackage{dvipsnames}{xcolor}

\usepackage{graphicx}
\usepackage{textcomp}
\usepackage{xcolor}
\usepackage{algorithm}
\usepackage[noend]{algpseudocode}
\usepackage{mathrsfs}
\usepackage{dsfont}

\usepackage{amsmath,amsfonts,amssymb,amsthm}
\usepackage{placeins}
\usepackage{subcaption}
\usepackage{blindtext}
\usepackage{tabularx}
\usepackage{a4wide}
\usepackage{caption}

\usepackage[T1]{fontenc}
\usepackage{hyperref}
\usepackage{tikz}
\usepackage{enumitem}
\usepackage[utf8]{inputenc}

\usepackage{siunitx}
\newlength{\mywidth}

\newcommand{\LRMSE}{\ensuremath{\mbox{\textit{LRMSE}}}}
\newcommand{\polylog}{\ensuremath{\mbox{polylog}}}

\newcommand{\commentlines}[1]{}

\newcommand{\loc}{\langle}
\newcommand{\roc}{\rangle}

\begin{document}

 \setlength {\marginparwidth }{0cm}


\mainmatter              
\title{Strongly Connected Components in Stream Graphs: Computation and Experimentations}
\titlerunning{Connected Components in Stream Graphs}  
%


\author{Léo Rannou\inst{1}\inst{2} \and Clémence Magnien\inst{1} \and Matthieu Latapy\inst{1}}
\authorrunning{Rannou et al.} 
%
\tocauthor{Léo Rannou, Clémence Magnien and Matthieu Latapy}
\institute{Sorbonne Université, CNRS, LIP6, F-75005 Paris, France
\and
Thales SIX, Theresis,
1 av. Augustin Fresnel,
91120 Palaiseau, France}
\setlength{\mywidth}{.9\columnwidth}
\maketitle              
\begin{abstract}
Stream graphs model highly dynamic networks in which nodes and/or links arrive and/or leave over time. Strongly connected components in stream graphs were defined recently, but no algorithm was provided to compute them. We present here several solutions with polynomial time and space complexities, each with its own strengths and weaknesses. We provide an implementation and experimentally compare the algorithms in a wide variety of practical cases. In addition, we propose an approximation method that significantly reduces computation costs, and gives even more insight on the dataset.
\end{abstract}




Connected components are among the most important concepts of graph theory. They were recently generalized to stream graphs \cite{latapy_stream_2018}, a formal object that captures the dynamics of nodes and links over time. Unlike other generalizations available in the literature, these generalized connected components {\em partition} the set of temporal nodes. This means that each node at each time instant is in one and only one connected component. This makes these generalized connected components particularly appealing to capture important features of objects modeled by stream graphs. However, computation of connected components in stream graphs has not been explored yet. Therefore, up to this date, they remain a formal object with no practical use. In addition, the algorithmic complexity of the problem is unknown, as well as the insight they may shed on real-world stream graphs of interest.

After introducing key notations and definitions (Section~\ref{sec-notations}), we present two algorithms for strongly connected components, together with their complexity (Section~\ref{sec-strongly}). We then apply these algorithms to several large-scale real-world datasets and demonstrate their ability to describe such datasets (Section~\ref{sec-experiments}). We also show that their performances may be improved greatly at the cost of reasonable approximations.

\section{The Stream Graph Framework}
\label{sec:sg}
\label{sec-notations}


Given any two sets $A$ and $B$, we denote by $A\otimes B$ the set of pairs $ab$ such that $a\in A$, $b\in B$ and $a\not= b$. Couples are ordered, while pairs are unordered: $(a,b) \neq (b,a)$ while $ab = ba$.

A \textbf{stream graph} $S = (T,V,W,E)$ is defined~\cite{latapy_stream_2018} by a finite set of nodes $V$, a time interval $T \subseteq\mathbb{R}$, a set of temporal nodes $W \subseteq T \times V$, and a set of links $E \subseteq T \times V \otimes V$ such that $(t,uv) \in E$ implies $(t,u) \in W$ and $(t,v) \in W$.

For any $u$ and $v$ in $V$, $T_u= \{t, (t,u) \in W\}$ denotes the set of time instants at which $u$ is present, and $T_{uv} = \{t, (t,uv) \in E\}$ the set of time instants at which $u$ and $v$ are linked together. We assume that both $T_u$ and $T_{uv}$ are unions of a finite number of disjoint closed intervals (possibly singletons) of $T$.

\begin{figure}[!ht]
\footnotesize
\centering
\includegraphics[width=0.5\mywidth]{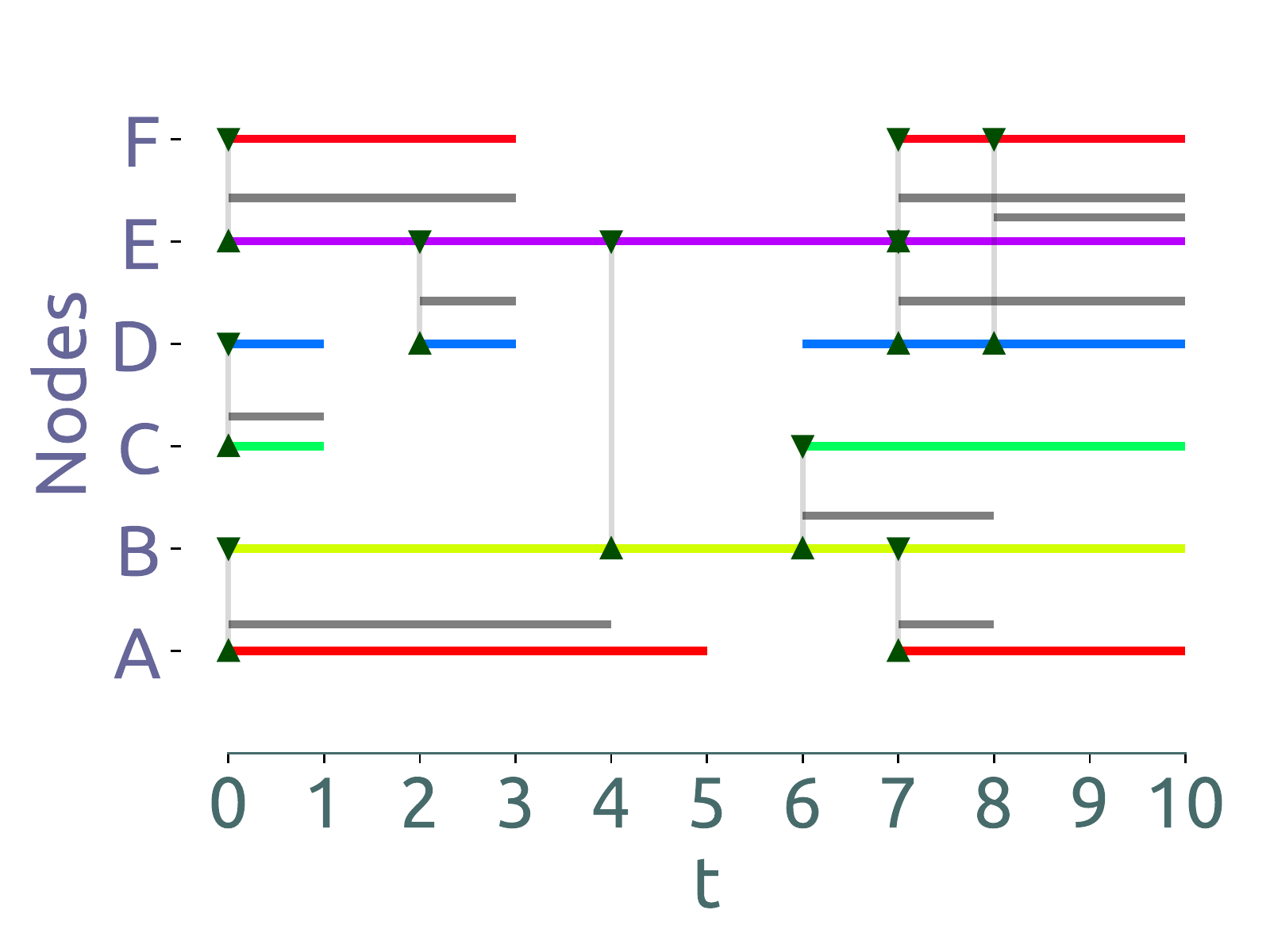}
\hfill
\includegraphics[width=0.49\mywidth]{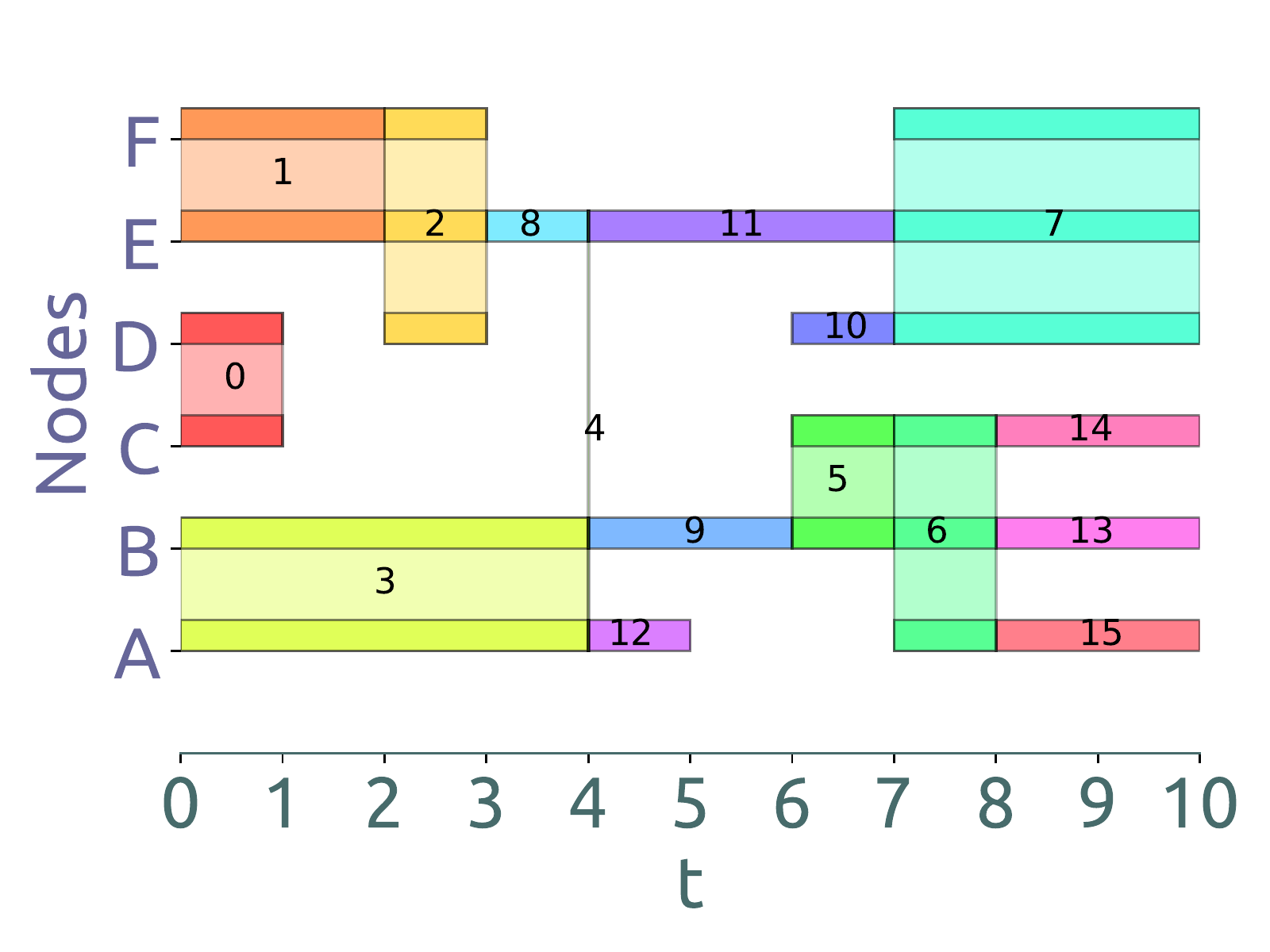}
\caption{
\footnotesize
(Left) An example of stream graph. We display time $T = [0,10]$ on the horizontal axis and nodes $V = \{A,B,C,D,E,F\}$ on the vertical one. We represent each node segment by a colored horizontal segment, with one color per node; and each link segment in grey by a vertical line between the two involved nodes at the link segment starting time, and an horizontal line from this time to its ending time.
(Right) The 16 strongly connected components of the stream graph.}
\label{fig:stream_graph_example}
\label{fig:stream_graph_scc}
\end{figure}

%

We call \textit{node segment} a couple $([b,e],u)$ such that $[b,e]$ is a segment that is not included in any other segment of $T_u$, and we denote by $\overline{W}$ the set of all node segments in $W$. We say that $b$ is an {\em arrival} of $u$, and $e$ a \textit{departure}. We denote by $N = |\overline{W}|$ the number of node segments in the stream. Likewise, we call \textit{link segment} a couple $([b,e],uv)$ such that $[b,e]$ is not included in any other segment of $T_{uv}$, and by $\overline{E}$ the set of all link segments in $E$. We say that $b$ is an {\em arrival} of $uv$, and $e$ a {\em departure}. We denote by $M = |\overline{E}|$ the number of link segments in the stream. We call all time instants that correspond to a node or link arrival or departure an {\em event time}. There are at most $2\cdot N + 2 \cdot M$ event times. Notice that the intervals considered above may be singletons. Then, $b=e$ and $[b,e] = \{b\} = \{e\}$. See Figure~\ref{fig:stream_graph_example} for an illustration.

The induced graph $G(S) = (V(S),E(S))$ is defined by  $V(S) = \{v, T_v \neq \emptyset \}$ and $E(S) = \{uv, \exists t, (t,uv)\in E\}$. We denote by $n = |V(S)|$ and $m = |E(S)|$ its number of nodes and links, respectively. We denote by $G_t = (V_t,E_t)$ the graph such that $V_t = \{v, (t,v)\in W\}$ and $E_t = \{uv, T_{uv} \neq \emptyset \}$. We denote by $G^-_t$ the graph that corresponds to the nodes and links present between the event time just before $t$ and $t$: $G^-_t = (V^-_t,E^-_t)$ where $V^-_t = \{v, \exists t' \neq t, [t',t] \subseteq T_v\}$ and $E^-_t = \{uv, \exists t' \neq t, [t',t] \subseteq T_{uv}\}$.




We consider in input a time-ordered sequence of node or link arrivals or departures. We maintain the set of present nodes and links at the current time instant $t$, {\em i.e.} the graph $G_t$, and we store their latest arrival time seen so far. This has a $\Theta(N+M)$ time and $\Theta(n+m)$ space cost for the whole processing of input data. Therefore, these worst-case complexities are lower bounds for our algorithms.

\section{Strongly Connected Components}
\label{sec-strongly}

A \textbf{strongly connected component} of $S=(T,V,W,E)$ is a maximal subset $I \times X$ of $W$ such that $I$ is an interval of $T$ and $X$ is a connected component of $G_t$ for all $t$ in $I$. It is denoted by $(I,X)$.
The set of all strongly connected components of $S$ is a partition of $W$~\cite{latapy_stream_2018}. See Figure~\ref{fig:stream_graph_scc} for an illustration.



Notice that some component time intervals are closed, some are open and some are a combination of the two. For instance, $([0,1],\{C,D\})$ is a closed component, $(]4,6[,\{B\})$ is an open one, $([6,7[, \{B,C\})$ is a left-closed and right-open one, and $([4,4], \{A,B,C\})$ is a closed and instantaneous component. Since the time intervals of components may be open or closed, we introduce the notation $\loc b, e \roc$ to indicate an interval that can be either open or closed on its extremities.
This interval contains $]b,e[$ and may or may not contain $b$ and/or $e$.
We will also use mixed notation: $\loc b, e]$ for instance designates an interval that may or may not contain $b$, but does contain $e$.


The number of strongly connected components is in $\Theta(N+M)$, because there can be one component per node segment, and each link segment may induce up to four components. Indeed, each beginning of a link segment may correspond to the beginning of two components: one instantaneous at the link segment beginning and one that starts just after; and each link segment ending may correspond to the beginning of two connected components if the corresponding component becomes disconnected. Explicitly writing a component to the output is done in linear time with respect to its number of nodes, in $\Omega(N+n\cdot M)$.




\subsection{Direct Approach}

One may compute strongly connected components directly from their definition, by processing event times in increasing order and by maintaining the set of strongly connected components that begin before or at current event time, and end after it.
We represent each such component as a couple $(\loc{} b, C)$,
meaning that it starts at $b$ (included or not) and involves nodes in $C$.

More precisely, we start with a set $\mathscr{C}$ containing $([\alpha,C)$ for each connected component $C$ of the graph $G_{\alpha}$ at the first event time $\alpha$. Then, for each event time $t>\alpha$ in increasing order we consider the connected components of $G^-_t$.
For each such component $C$, if there is no component $(\loc b,X)$ with $X=C$ in $\mathscr{C}$
then we add $(]t',C)$ to $\mathscr{C}$, where $t'$ is the event time preceding $t$.
For each element $(\loc b,X)$ of $\mathscr{C}$, if $X$ is not a connected component of $G^-_t$, then we remove it from $\mathscr{C}$ and we output $(\loc b,t'],X)$.
We then turn to the connected components of $G_t$:
for each such component $C$, if there is no 
component $(\loc b,X)$ with $X=C$ in $\mathscr{C}$
then we add $([t,C)$ to $\mathscr{C}$; and for each element $(\loc b,X)$ of $\mathscr{C}$, if $X$ is not a connected component of $G_t$, then we remove it from $\mathscr{C}$ and we output $(\loc b,t[,X)$.
Finally, when the last event time $t=\omega$ is reached, we output $(\loc b,\omega],X)$ for each element $(\loc b,X)$ of $\mathscr{C}$.

Clearly, this algorithm outputs all strongly connected components of the considered stream graph. Computing the connected components of each graph is in $O(n+m)$ time and space. The considered set families (the graph connected components, as well as the elements of $\mathscr{C}$) form partitions of $V$. Therefore, their storage and all set comparisons processed for each event time have a cost in $O(n)$ time and space. There are $O(M+N)$ event times, therefore, the time complexity of this method is $O((N+M) \cdot (n+m))$, and it needs $O(n+m)$ space.

Without changing its time complexity, this algorithm may be improved by ignoring event times $t$ such that all events occurring at $t$ are link arrivals between nodes already in the same connected component. However, one still has to compute graph connected components at each event time with link departures. Therefore, this improvement is mostly appealing if many link departures occur at the same event times.

More generally, the approach above is efficient only if many events (node and/or links arrivals and/or departures) occur at each event time. Then, many connected components may change at each event time, and computing them from scratch makes sense. Instead, if only few events occur at most event times, managing each event itself and updating current connected components accordingly is appealing.

This leads to the following algorithm, which starts with an empty set $\mathscr{C}$, considers each event time $t$ in increasing order, and performs the following operations.
\begin{enumerate}
    \item
    \label{alg:naive:nadd}
    For each node segment $([b,e],u)$ such that $b = t$ (node arrival), add $([b,\{u\})$ to $\mathscr{C}$.
    \item
    \label{alg:naive:ladd}
    For each link segment $([b,e],uv)$ such that $b = t$ (link arrival), let $C_u = (\loc b_u,X_u)$ and $C_v = (\loc b_v,X_v)$ be the elements of $\mathscr{C}$ such that $u\in X_u$ and $v\in X_v$; 
    if $C_u \neq C_v$ then replace $C_u$ and $C_v$ by $([t,X_u \cup X_v)$ in $\mathscr{C}$.
    Then: if $\loc{} b_u \not = [t$ then output $(\loc b_u,t[,X_u)$;
    if $\loc{} b_v \not= [t$  then output $(\loc b_v,t[,X_v)$.
    \item \label{alg:naive:lremoval}
    Let $G'_t = G_t$; then
    for each link segment $([b,e],uv)$ such that $e = t$ (link departure), let $C_v = C_u = (\loc b_u,X_u)$ be the element of $\mathscr{C}$ such that $u\in X_u$ and $v\in X_u$;
    remove the link $uv$ from $G'_t$;
    if there is no path between $u$ and $v$ in $G_t'$ 
    then replace $C_u$ by  $C_u' = (]t,X_u')$ and $C_v' = (]t,X_v')$ in $\mathscr{C}$ where $X_u'$ and $X'_v$ are the connected components of $u$ and $v$ in $G_t'$, respectively;
    if $\loc{} b_u \not= ]t$ then output $(\loc b_u,t],X_u)$.
    \item \label{alg:naive:nremoval}
    For each node segment $([b,e],u)$ such that $e = t$ (node departure),
    let $C_u = (\loc b_u,X_u)$ be the element of $\mathscr{C}$ such that $u\in X_u$;
    remove $C_u$ from $\mathscr{C}$;
    if $\loc b_u \neq ]t$ then output $(\loc b_u,t],\{u\})$.
\end{enumerate}

We call this algorithm {\em SCC Direct}. It clearly outputs the strongly connected components of the considered stream, like the previous algorithm. It performs $2(M+N)$ of the steps above, corresponding to $N$ node arrivals and departures and $M$ link arrivals and departures. One easily deals with node arrivals and departures in constant time. If a link arrival induces a merge between two components, computing their union is in $O(n)$, as is outputting both components if needed. Thus the complexity for link arrival steps is in $O(M\cdot n)$.
Each link departure calls for a computation of the connected components of a graph, and writing a component to the output is in $O(n)$. Thus the complexity for link departure steps is in $O(M\cdot(m+n))$. We obtain a total time complexity in $O(M\cdot(m+n)+N)$. The space complexity is still in $O(n+m)$ as above.


\subsection{Fully Dynamic Approach}


The SCC Direct algorithm presented above is strongly related to one of the most classical algorithmic problems in dynamic graph theory, called fully dynamic connectivity \cite{kejlberg_faster_2016,wulff_faster_2013,iyer_experimental_2001,alberts_empirical_1997,huang_fully_2017,kapron_dynamic_2013,henzinger_randomized_1999}, which aims at maintaining the connected components of an evolving graph. Considering a sequence of link additions and removals, dynamic connectivity algorithms maintain a data structure able to tell if two nodes are in the same connected components ({\em query} operation) and to merge or split connected components upon link addition or removal ({\em update} operation).

This data structure and the corresponding operations can be used in the above algorithm:
we can use the data structure to store $\mathscr{C}$, the set of current connected components (we also need to store the beginning time of each component, which has negligible cost).
Then, at each link arrival or departure, we can use the query operation to test whether the two nodes are in the same component or not, 
and the update operation to add or remove the current link to the data structure, while keeping an up-to-date set of connected components.
When we observe a node appearance it is necessarily isolated, so we have to add the current time to its component.
All the other steps (mainly, writing the output) are unchanged. We call this algorithm {\em SCC FD}.

Several methods efficiently solve the dynamic connectivity problem, the key challenge being to know if updates and queries may be performed in $O(\log(n))$ time, where $n$ is the number of nodes in the graph. Current exact solutions perform updates in $O\left(\sqrt{\frac{n\cdot(\log\log(n))^2}{\log(n)}}\right)$ worst time \cite{kejlberg_faster_2016}, or in $\frac{\log^2(n)}{\log\log(n)}$ amortized worst time \cite{wulff_faster_2013}. Probabilistic (exact or approximate) methods perform even better, but they remain above the $O(\log(n))$ time cost \cite{huang_fully_2017,kapron_dynamic_2013,henzinger_randomized_1999}.

It is well acknowledged that these algorithmic time and space complexities hide big constants, and that the underlying algorithms and data structures are very intricate. As a consequence, implementing these algorithms is an important challenge in itself \cite{alberts_empirical_1997,iyer_experimental_2001}, and the results above should be considered as theoretical bounds. In practice, the implemented algorithms typically have $O(\log(n)^3)$ amortized time and linear space complexities, still with large constants \cite{alberts_empirical_1997,iyer_experimental_2001}.

In SCC FD, we perform $O(M)$ updates and queries, which leads to a $O(M\cdot\polylog(n))$ overall time cost for these operations, with any of the polylog dynamic connectivity algorithms cited above. This cost is dominated by the cost of outputting the results, which is in $O(M\cdot n)$. 
An additional $N$ factor is needed to deal with node arrivals and departures. Hence, we obtain a total time in $O(M\cdot n + N)$.
The space cost of dynamic connectivity methods is in $O(m + n\cdot\log{} n)$, and we do not store significantly more information.


This algorithm is particularly appealing if large connected components are quite stable, {\em i.e.} if most largest strongly connected components in the stream have a long duration. Indeed, in this case, fully dynamic update operations are much faster than updates used in SCC Direct, and the output is much smaller than the maximum $\Omega(N+M\cdot n)$ bound. The cost of SCC FD is then dominated by fully dynamic operations, and its time complexity is reduced to $O(M\cdot \polylog(n))$.

\section{Experiments and Applications}
\label{sec-experiments}



In this section, we conduct thorough experiments with several real-world datasets and our different algorithms. SCC Direct was significantly faster, and only SCC Direct was able to perform the computation in central memory of large-scale datasets (several dozens of millions of link segments). We publicly provide Python 3 implementations of our algorithms in the Straph library~\cite{Straph}.


\subsection{Datasets}
\label{subsec-datasets}


First notice that most available datasets record instantaneous interactions only, either because of periodic measurements, or because only one timestamp is available. In such situations, one resorts to $\delta$-analysis~\cite{latapy_stream_2018}: one considers that each interaction lasts for a given duration $\delta$. This transforms a dataset into a stream graph $S = (T,V,W,E)$ in which all link segments last for at least $\delta$, and all links in $D$ separated by a delay lower than $\delta$ lead to a unique link segment. Nodes are considered as present only when they have at least one link.

In order to explore the performances of our algorithms in a wide variety of situations, we considered 14 publicly available datasets that we shortly present below. Their key stream graph properties are given in Table~\ref{tab:datasets_description_size}, together with the value of $\delta$ we used. It either corresponds to a natural value underlying the dataset or is determined by the original timestamp precision.

\begin{table}[!ht]
\footnotesize
\centering
\resizebox{0.5\textwidth}{!}{\begin{tabular}{|ll|l|l|l|l|l|l| } 
	 	  \hline
	 	 & $\delta$ & $n$ & $m$ & $|T|$ & $N$ & $M$ \\ 
	 	 \hline 
	 	 UC & 1h & 2K & 14K & 189d & 43K & 34K  \\ 
	 	 \hline 
	 	 HS 2012 & 60s & 327 & 6K & 4d  & 48K & 46K \\
	 	\hline
	 	 Digg & 1h & 30K & 85K & 14.5d  & 110K & 86K \\
	 	\hline
	 	Infectious & 60s & 11K & 45K & 80d  & 85K & 133K \\
	 	\hline 
	 	Twitter  & 600s & 304K & 452K & 7d & 543K & 488K \\
	    \hline 
	 	 Linux & 10h & 27K & 160K & 8y & 450K & 544K  \\
	     \hline 
	 	 Facebook & 10h & 46K & 183K & 4.3y & 957K & 588K  \\ 
	 	\hline
	 	Epinions & 10h & 132K & 711K & 2.6y & 404K & 743 K \\
	 	\hline
	 	Amazon & 1h & 2.1M & 5.7M & 9.5y  & 9.9M & 5.8M \\
	 	\hline
	 	Youtube & 24h & 3.2 M & 9.4M & 226d & 6.7M & 9.4M \\
	 	\hline
	 	Movielens & 1h & 70K & 10M & 14y  & 8.5M & 10M \\
	 	\hline 
	 	 Wiki & 1h & 2.9M & 8.1M & 14.3y  & 18.3M & 14.5M  \\ 
	 	 \hline 
	 	 Mawilab & 2s & 940K & 9.1M & 902s  & 17M & 18.8M \\ 
	 	\hline 
	 	Stackoverflow & 10h & 2.6M & 28.2M & 7.6y & 30M & 33.5 M \\
	 	\hline
	 \end{tabular}}
\medskip
\caption{
\footnotesize
Key features of the real-world stream graphs we consider, ordered with respect to their number $M$ of link segments (K indicates thousands, M millions).}
\label{tab:datasets_description_size}
\end{table}

\textbf{UC Message (UC)}~\cite{kunegis2013konect} is a capture of messages between University of California students in an online community.
\textbf{High School 2012 (HS 2012)}~\cite{high_school_2012} is a sensor recording of contacts between students of 5 classes during 7 days in a high school in Marseille, France in 2012.
\textbf{Digg}~\cite{kunegis2013konect} is a set of links representing replies of Digg website users to others.
\textbf{Infectious}~\cite{infectious} is a recording of face-to-face contacts between visitors of an exhibition in 2009, Dublin.
\textbf{Twitter Higs (Twitter)}~\cite{twitter_higs,snapnets} is a recording of all kinds of twitter activity for one week around the discovery of the Higgs boson in 2012.
\textbf{Linux Kernel mailing list (Linux)}~\cite{kunegis2013konect} represents the email replies between users on this mailing-list.
\textbf{Facebook wall posts (Facebook)}~\cite{viswanath-2009-activity} represents messages exchanged between Facebook users, through their walls.
\textbf{Epinions}~\cite{kunegis2013konect} is a set of timestamped trust and distrust link creations on Epinions, an online product rating site.
\textbf{Amazon}~\cite{kunegis2013konect} contains product ratings on Amazon.
\textbf{Youtube}~\cite{youtube} is a social network of YouTube users and their friendship connections.
\textbf{Movielens}~\cite{movielens} contains movie ratings  by users of the Movielens site.
\textbf{Wiki Talk En (Wiki)}~\cite{kunegis2013konect} is a recording of discussions between contributors to the English Wikipedia.
\textbf{Mawilab 2020-03-09 (Mawilab)}~\cite{mawilab} is a 15 minutes capture of network traffic on a backbone trans-pacific router in Japan on March 3, 2020. Each link represents a packet exchanged between two internet addresses.
\textbf{Stackoverflow}~\cite{stackoverflow,snapnets} is a recording of interactions on the stack overflow web site.

\subsection{Algorithm performances}

\commentlines{
\footnotesize
\begin{table}[!ht]
\centering
\begin{tabularx}{\mywidth}[t]{|c|X|X||X|X|X|} 
	 	  \hline
	 	 & $|\mathscr{W}|$ & $|\mathscr{C}|$ & \textit{WCC UF} & \textit{SCC Direct} & \textit{SCC FD}\\ 
	 	 \hline 
	 	 UC & 11K & 54K & 0.35 & \textbf{0.47} & 6.83  \\
	 	 \hline 
	 	 HS 2012 & 13K & 50K & 0.45 & \textbf{0.41} & 5.11 \\
	 	 \hline
	 	 Digg & 26K & 144K & 1.0 & \textbf{1.34} & 19.4 \\
	 	\hline 
	 	Infectious & 16.9K & 106.3K & 0.80 & \textbf{1.18} & 16.0  \\
	 	 \hline 
	 	Twitter & 113K & 580K & 3.94 & \textbf{33.89} & 9.98K \\
	 	\hline 
	 	Linux & 63K & 698K & 3.8 & \textbf{11.6} & 227.0  \\ 
	 	 \hline 
	 	Facebook & 373K & 795.0K & 10.36 &  \textbf{7.96} & 55.3  \\ 
	 	\hline
	 	Epinions & 75.6K & 75.6K & 5.6 & \textbf{3.3} & - \\
	    \hline
	 	Amazon & 4.1M & 4.1M & 117.3 & \textbf{102.3} & 691.1 \\
	    \hline
	 	Youtube & 738.6K & 1.2M & 96.2 & \textbf{150.8} & - \\
	 	\hline
	 	Movielens & 282K & 12.5M & 93.3 & \textbf{340.0} & - \\
	 	\hline 
	 	Wiki & 4.2M & 23.5M & 247.34 & \textbf{247} & -  \\ 
	 	\hline 
	 	Mawilab & 1.1M & 30M & 430 &\textbf{90K} & - \\
	 	\hline
	 	Stackoverflow & 5.2M & 45.9M & 410.7 & \textbf{36.7K} & - \\
	 	\hline
	 \end{tabularx}
\caption{
\footnotesize
Number of WCC ($|\mathscr{W}|$) - Number of SCC ($|\mathscr{C}|$) - Algorithms running time in seconds (K = $10^3$, M=$10^6$)}
\label{tab:scc_algorithms}
\end{table}
}

\begin{figure}[!ht]
\footnotesize
\centering
\includegraphics[width=0.7\textwidth]{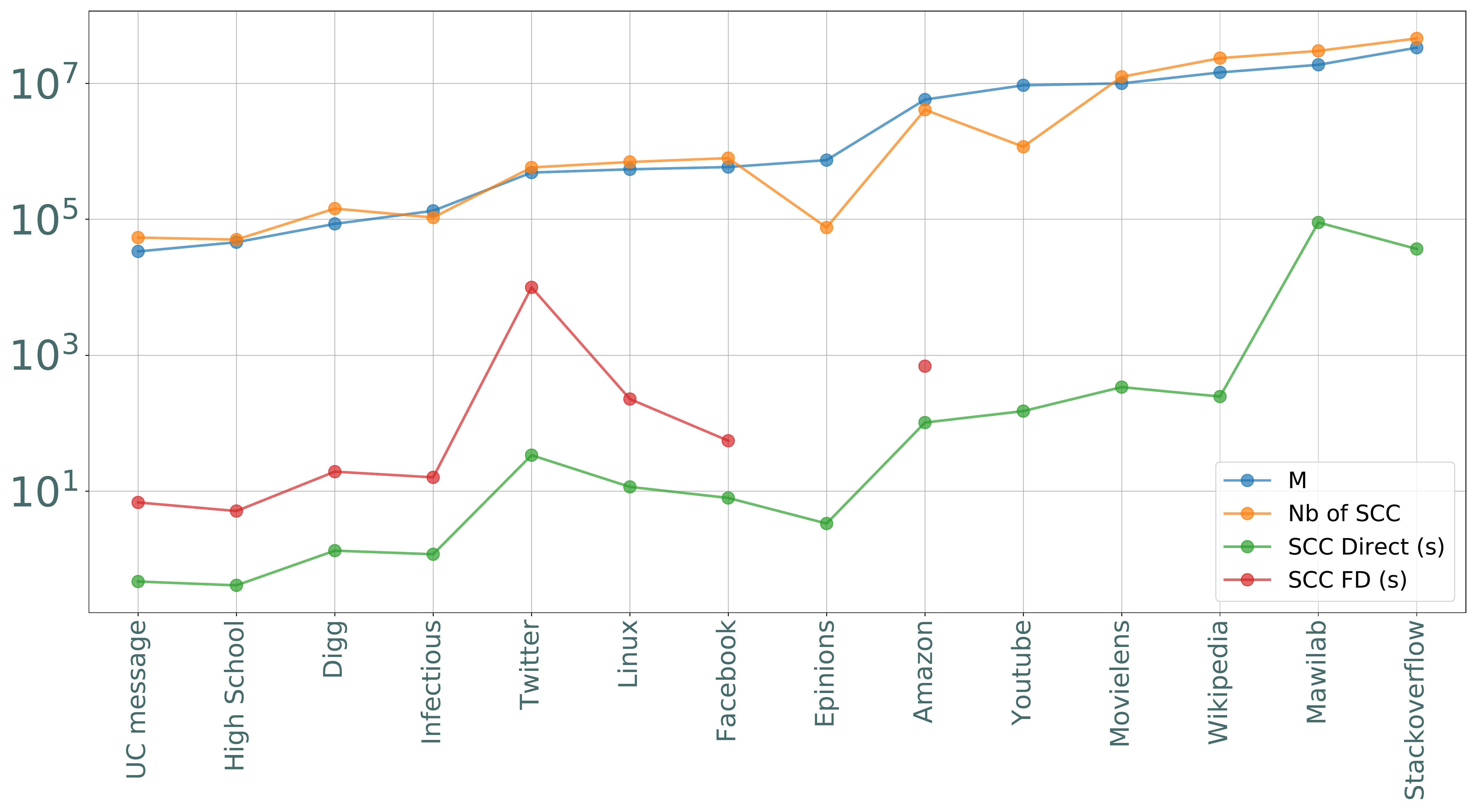}
\caption{
\footnotesize
Time cost of \textit{SCC Direct} and \textit{SCC FD} in seconds, along with the number $M$ of link segments and the number of strongly connected components, for each considered stream (horizontal axis, ordered with respect to $M$).
}
\label{fig:mawi_running_time_vs_event_times}
\label{fig:mawi_correlations}

\end{figure}




Figure~\ref{fig:mawi_running_time_vs_event_times} presents the time cost for each dataset, and show a strong relation between the number of link segments, the number of connected components, and computation time. Notice however that \textit{Wiki} and \textit{Mawilab} have similar numbers of link and node segments but \textit{SCC Direct} is several order of magnitude faster on \textit{Wiki}. This difference comes from their quite different structure regarding connected components: \textit{Mawilab} has more than $21M$ SCC involving at least $30K$ nodes, whereas \textit{Wiki} has only $2K$ such SCC. As explained in Section~\ref{sec-strongly}, the computational cost of \textit{SCC Direct} mainly depends on the number of nodes in each SCC, which is observed in this experiment.



\subsection{Connectedness analysis of IP traffic}

\begin{figure*}[!ht]
\footnotesize
\centering
\includegraphics[width=0.32\textwidth]{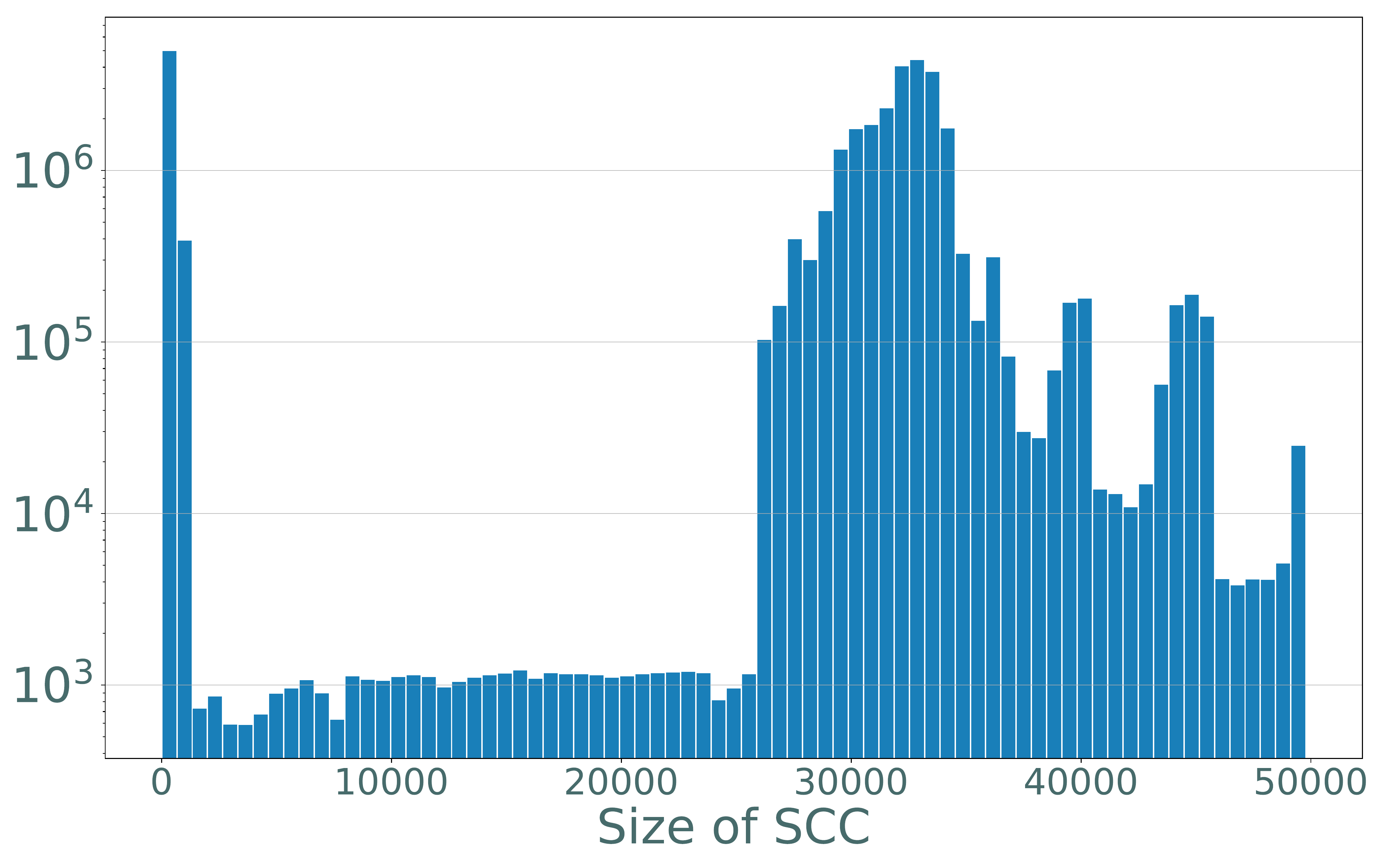}
\hfill
\includegraphics[width=0.32\textwidth]{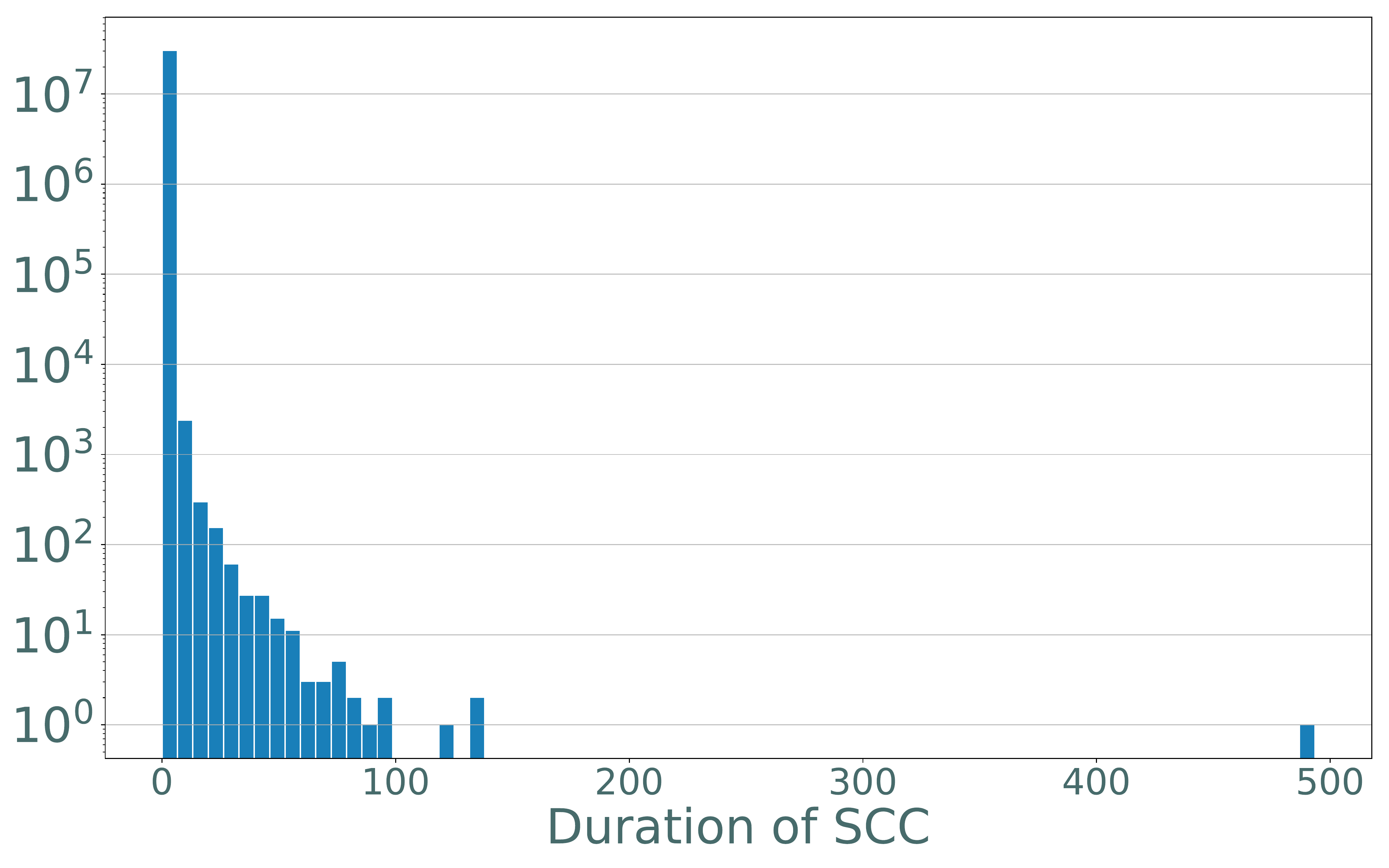}
\hfill
\includegraphics[width=0.3\textwidth]{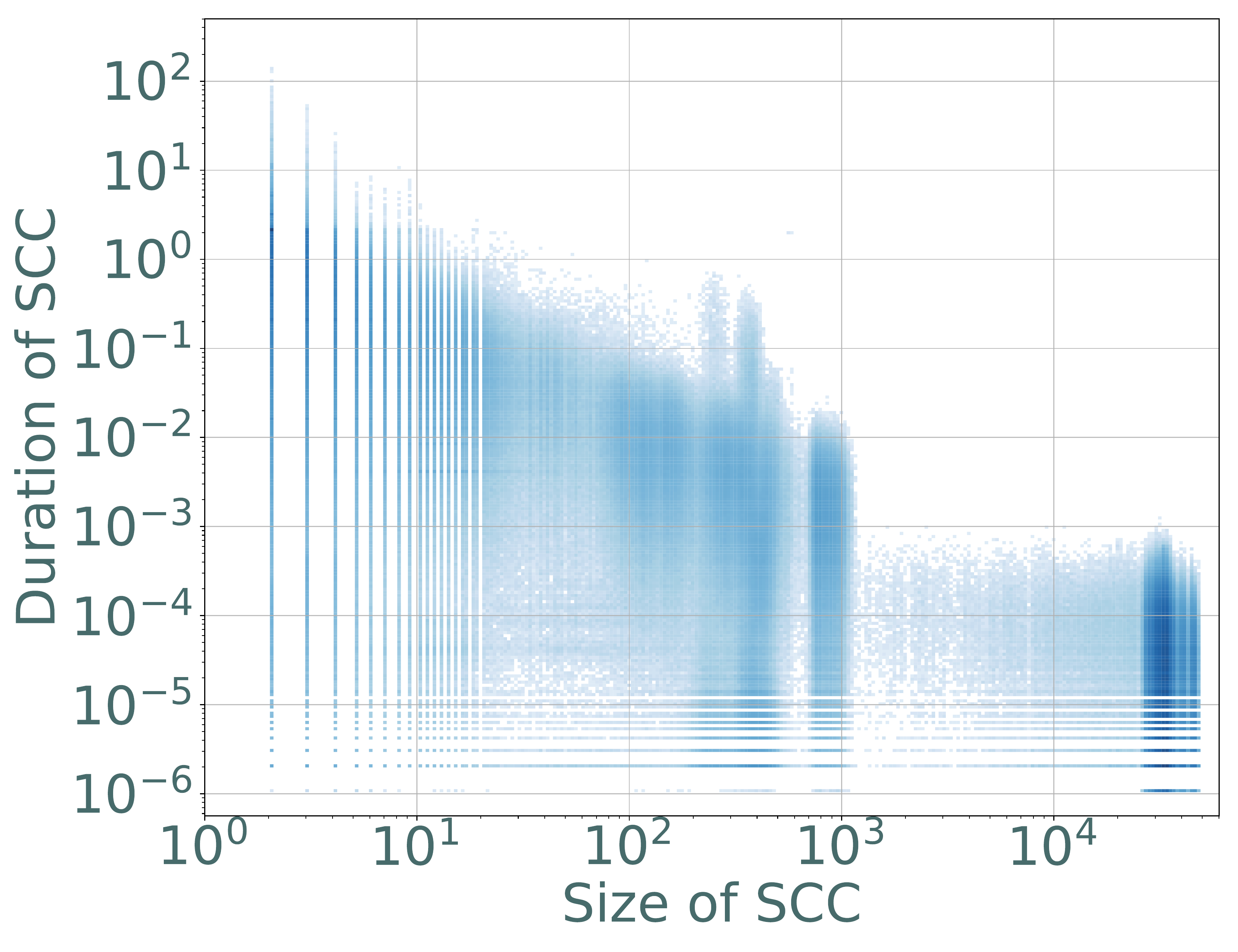}
\caption{
\footnotesize
Distribution of the size (left) and duration (middle) of SCC in \textit{Mawilab} dataset. Duration of each SCC as a function of its size, in log-log scales (right).}
\label{fig:mawi_size_vs_duration_scc}
\end{figure*}

We take the \textit{MawiLab} IP traffic capture as a typical instance of large real-world datasets modeled by a stream graph, and we use it to illustrate the relevance of connected component analysis. The stream has $30,062,184$ such components, with no giant one. Given $C=(\loc b,e \roc,X)$ we call the number of involved nodes $|X|$ its size, and the length of its time interval $e-b$ its duration. 

In Figure~\ref{fig:mawi_size_vs_duration_scc}, we display the strongly connected component size and duration distributions as well as the duration of components as a function of their size. Clearly, component size and duration are not linearly dependant. \textbf{No component significantly stands out of the crowd: there is no component with both a long duration and a large size}. Instead, large strongly connected components have a very short duration, and, conversely, long components have a small size. For instance, all components involving at least $2K$ nodes have a duration lower than $\num{1e-3}$ seconds, and all components that last for more than two seconds involve less than ten nodes. The largest component (in terms of number of nodes) involves $49,791$ nodes (only $5.3$\% of the whole), and lasts for $\num{8.2e-5}$ seconds (only $\num{9.1e-6}$\% of the whole).

More generally, these plots show that there are many strongly connected components with very short duration: 90\% last less than $0.14$ seconds. These components are due to the \textit{frontier effect}, that we define as follows. Consider a set $X$ of nodes, and assume that link segments that start close to a given time $b$ and end close to a given time $e$ connect them. However, they all start at different times and end at different times. This leads to a connected component $([b',e'],X)$, with $b'$ close to $b$ and $e'$ close to $e$, but also to many short strongly connected components that both start and end close to $b$, or close to $e$. These components make little sense, if any, but they account for a huge fraction of all strongly connected components, and so they have a crucial impact on computation time as explained in the previous section. We show below how to get rid of them while keeping crucial information.

%
%

\subsection{Approximate Strongly Connected Components}
\label{sec-approximation}

The fact that link segments start and end at slightly different times induces many strongly connected components of very low duration, that have little interest. We, therefore, propose to consider the following approximation of the stream graph $S = (T,V,W,E)$.
Given an approximation parameter $\Delta < \delta$ and any time $t$ in $T$, we define $\lfloor t\rfloor_\Delta$ as $\Delta \cdot\lfloor \frac{t}{\Delta}\rfloor$ and $\lceil t\rceil_\Delta$ as $\Delta \cdot\lceil \frac{t}{\Delta}\rceil$. We then define $S_\Delta = (T,V,W_\Delta,E_\Delta)$ where $W_\Delta = \cup_{([b,e],v) \in \overline{W}} [\lceil b\rceil_\Delta,\lfloor e\rfloor_\Delta]\times\{v\}$ and $E_\Delta = \cup_{([b,e],uv) \in \overline{E}} [\lceil b\rceil_\Delta,\lfloor e\rfloor_\Delta]\times\{uv\}$.
In other words, we replace each node segment $([b,e],v)$ by a shorter node segment that starts at the first time after $b$ and ends at the last time before $e$ which are multiple of $\Delta$. We proceed similarly with link segments.

First notice that $S_\Delta$ is an approximation of $S$, in the sense that $S_\Delta$ may be computed from $S$, but not the converse. In addition, each node or link segment in $S$ lasts at least $\delta$, and since $\Delta$ is lower than $\delta$, no node or link segment disappears when $S$ is transformed into $S_\Delta$; only their starting and ending times change. In addition, $S_\Delta$ is included in $S$: $W_\Delta \subseteq W$ and $E_\Delta \subseteq E$. This has an important consequence: all paths in $S_\Delta$ are also paths in $S$, and so the approximation does not create any new reachability relation. It, therefore, preserves key information contained in the original stream.


\begin{figure}[!ht]
\footnotesize
\centering
\includegraphics[width=0.8\mywidth]{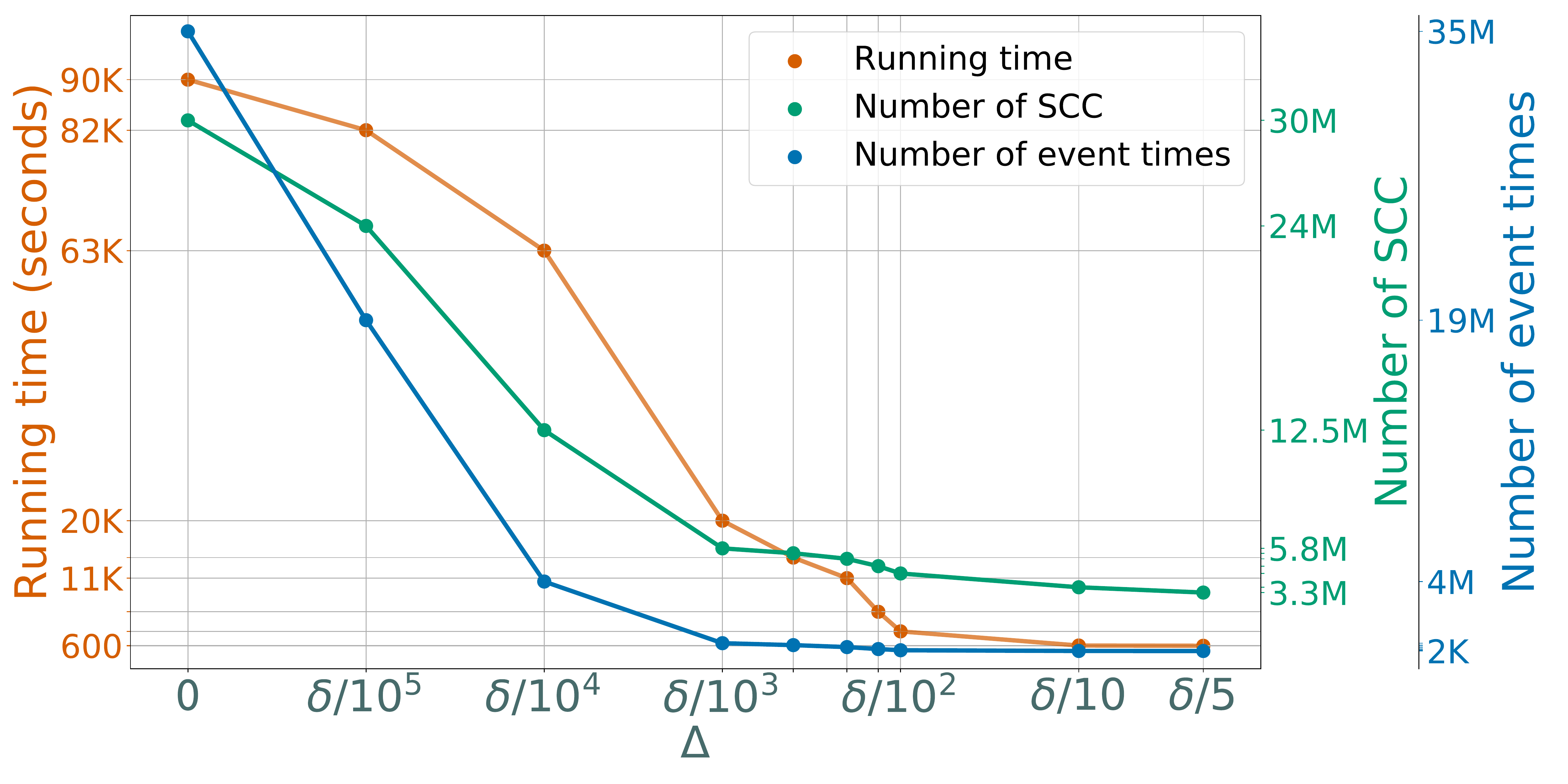}
\caption{
\footnotesize
Running time of \textit{SCC Direct}, number of SCC and number of event times in MawiLab, as a function of $\Delta$ (here, $\delta = 2s$).}
\label{fig:mawi_n_scc_according_to_delta}
\end{figure}

Let us first observe the effect of the approximation on strongly connected components in Figure~\ref{fig:mawi_n_scc_according_to_delta}. The number of components rapidly drops from its initial value of $30$ millions (for $\Delta=0$, {\em i.e.} no approximation) to less than $6$ millions for $\Delta=\delta/10^3= 0.002$. Its decrease is much slower when $\Delta$ grows further, which indicates that the stream does not anymore contain an important number of irrelevant components due to the \textit{frontier effect}. As expected, this has a strong impact on computation time, which we also display; it also very rapidly drops, from more than one day to less than one hour, making computations on such large-scale datasets much quicker.

\begin{figure*}[!ht]
\footnotesize
\centering
\includegraphics[width=.32\textwidth,scale=1.1]{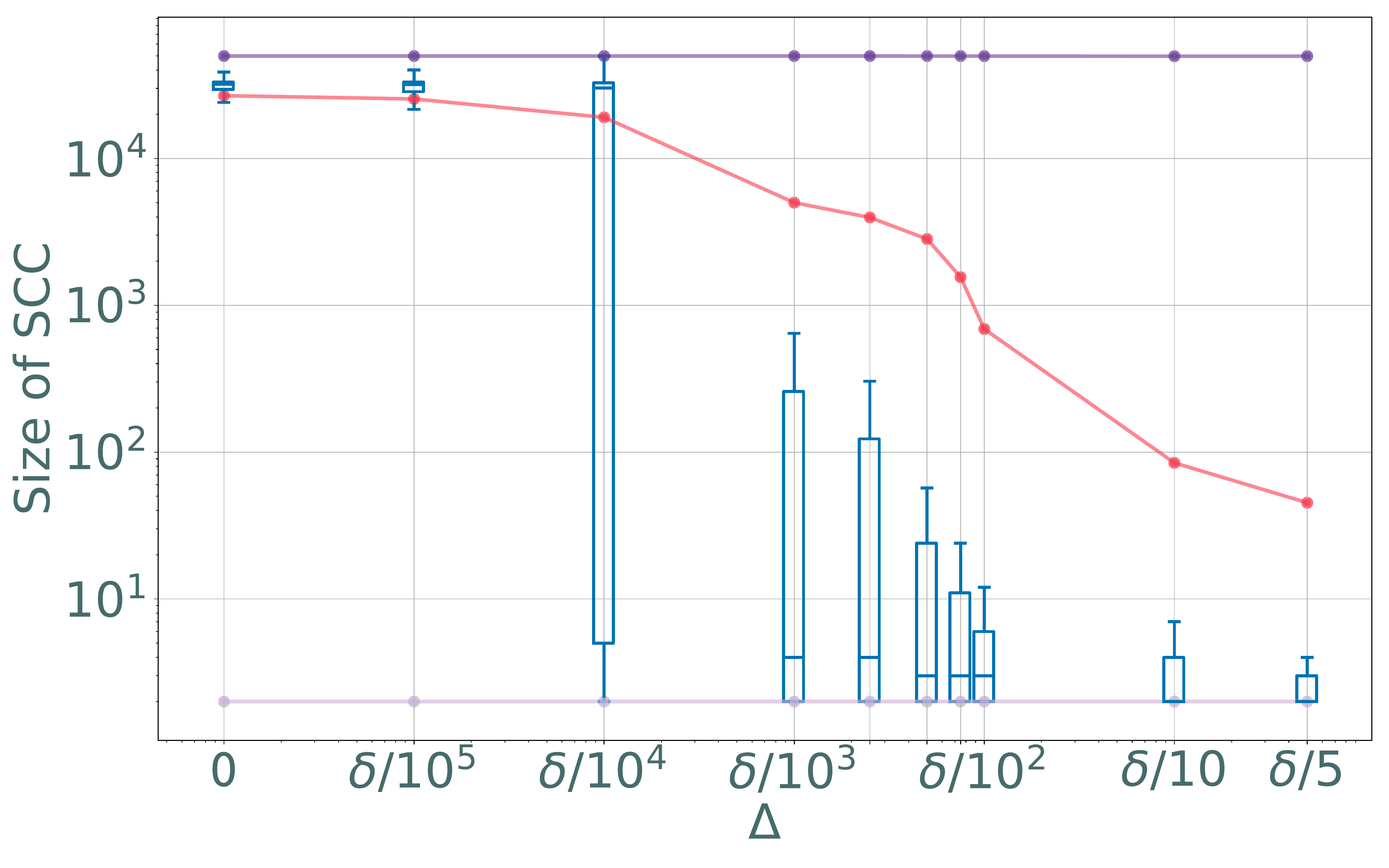}
\hfill
\includegraphics[width=.32\textwidth,scale=1.1]{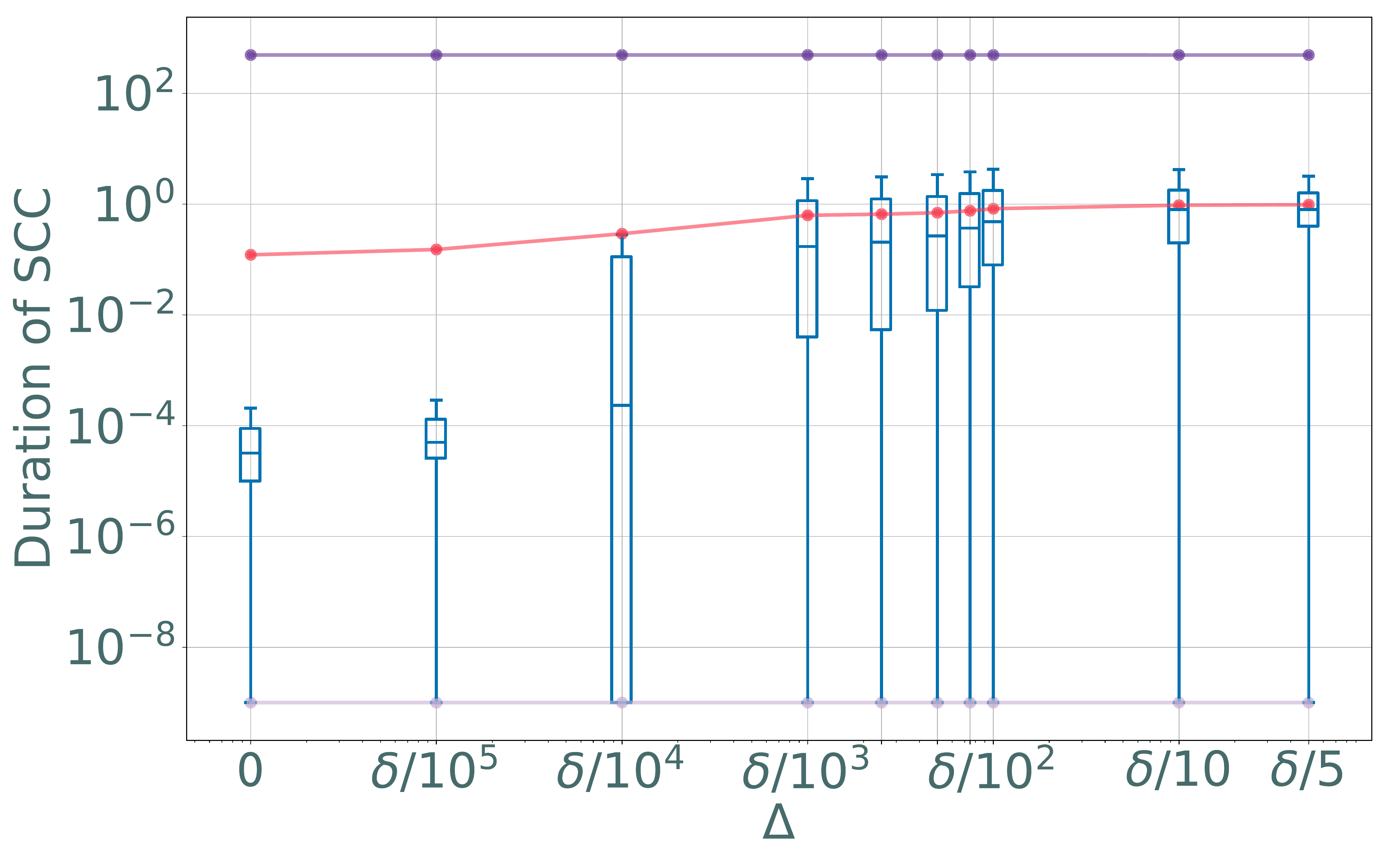}
\hfill
\includegraphics[width=.32\textwidth,scale=1.1]{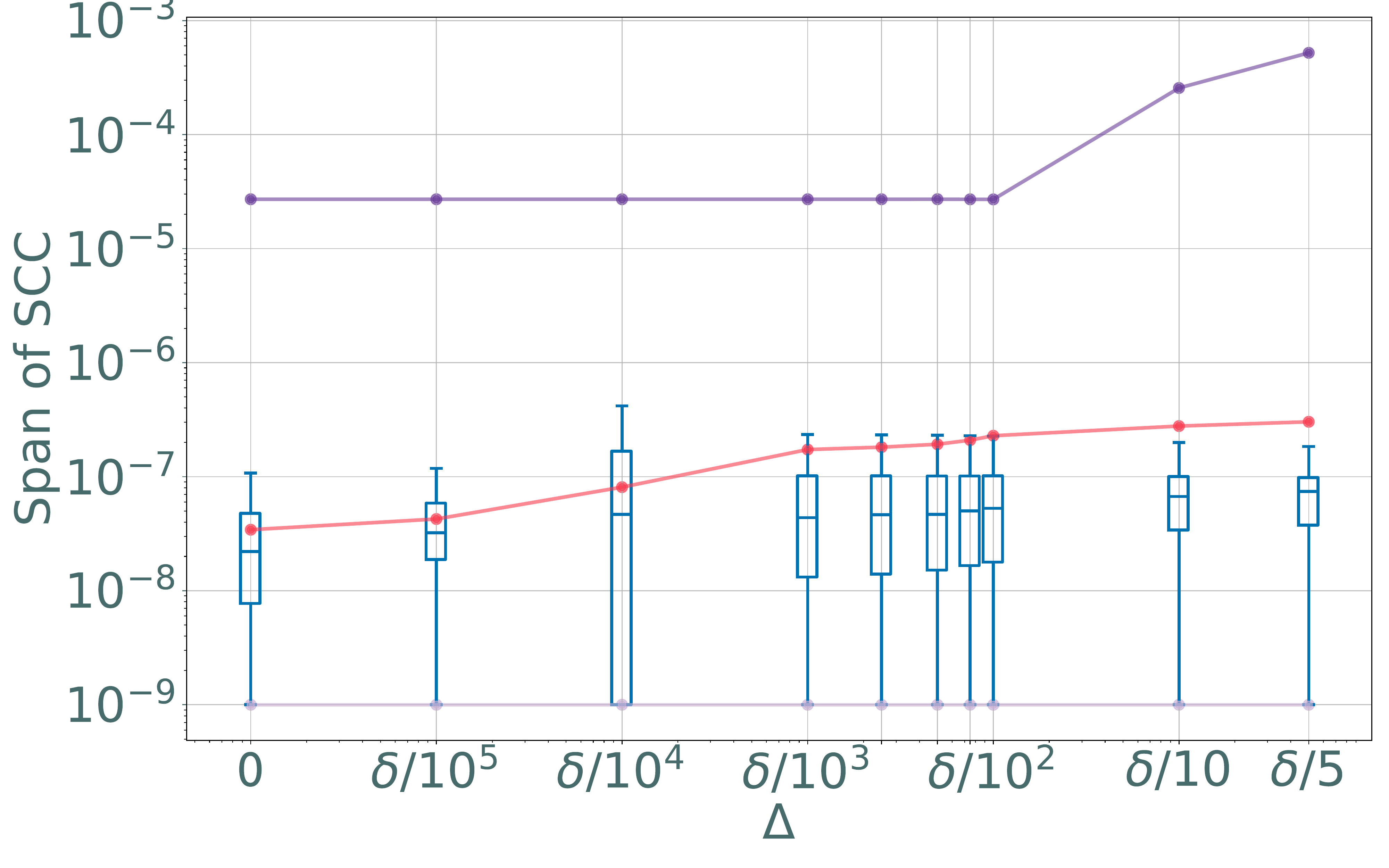}
\caption{
\footnotesize
Box plots representing the distribution of the size (left), duration (middle) and span (right) of strongly connected components in \textit{Mawilab}, for various values of $\Delta$ (here, $\delta = 2s$). We indicate the mean, minimal, and maximal values with dots connected by horizontal lines, as well as the median and percentiles with vertical boxes.}
\label{fig:mawi_boxplots_according_to_delta}
\end{figure*}

Figure~\ref{fig:mawi_boxplots_according_to_delta} presents the effect of $\Delta$ on size, duration and span distributions of strongly connected components.
For $\Delta = \delta/10^4$, we notice that while the number of components has decreased by half only fifty percent of them involve more than $30K$ nodes. Furthermore, as $\Delta$ increases, the number of components tends to be stable (Figure~\ref{fig:mawi_n_scc_according_to_delta}) but the number of components involving more than $30K$ nodes continues to drop. This explains the differences observed in the execution time of \textit{SCC Direct} (Figure~\ref{fig:mawi_n_scc_according_to_delta}) and confirms that the approximation eliminates most very short connected components, but not all: the ones which are not due to the frontier effect are preserved, another wanted feature.


\subsection{Application to Latency Approximation}

\begin{figure}[!ht]
\footnotesize
\centering
\includegraphics[width=.8\textwidth]{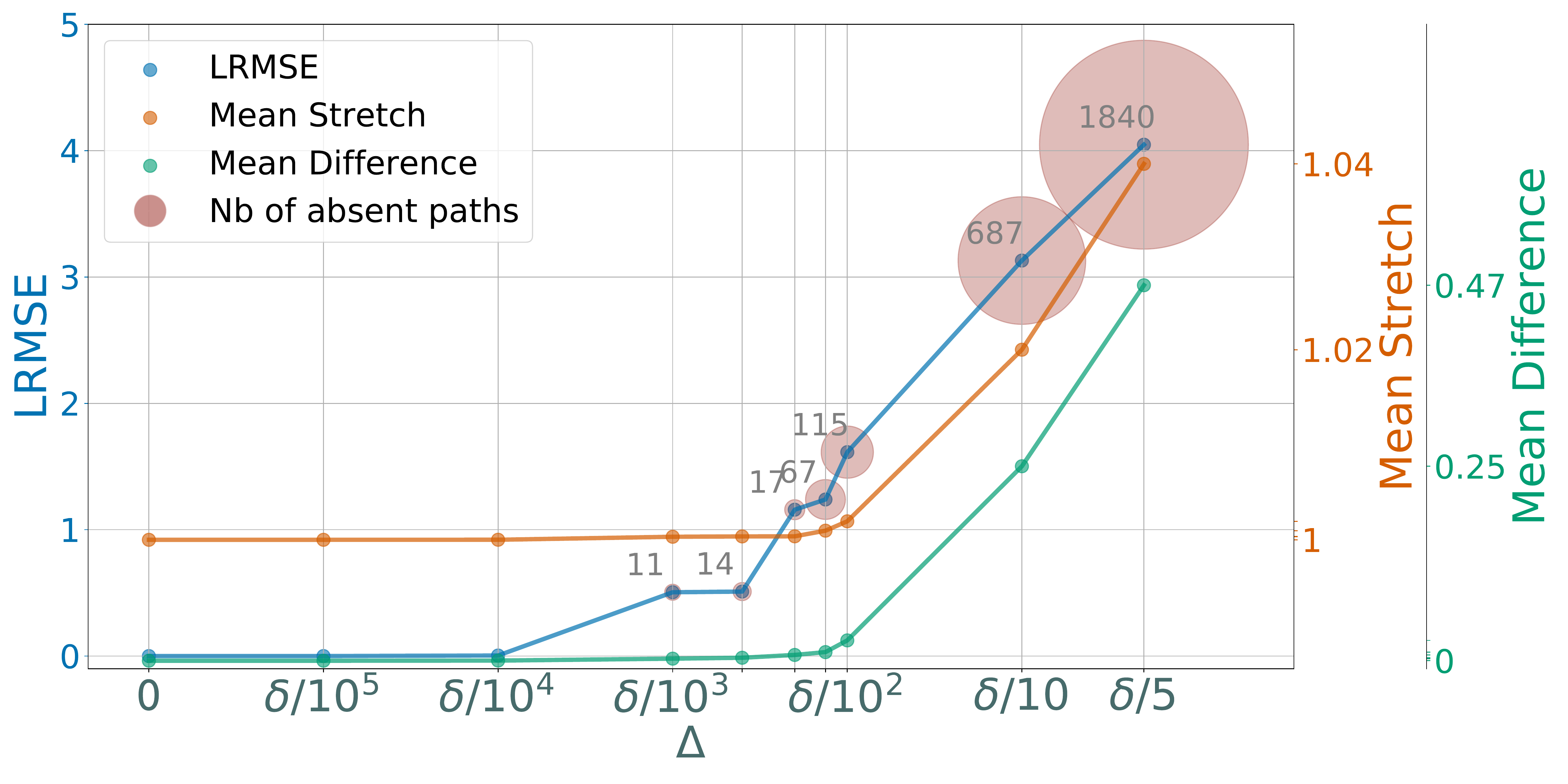}
\caption{
\footnotesize
Evolution of the LRMSE,  the average difference between latencies and the average latency stretch with respect to $\Delta$ in \textit{Mawilab}. We indicate the number of missing paths and represent it as a disk of area proportional to this number.}
\label{fig:mawi_lrmse_according_to_delta}
\end{figure}


Although the approximation above has a strong impact on the number of strongly connected components, it preserves key information of the stream. We illustrate this by considering one of the most widely studied features of these objects: the latency between nodes~\cite{kempe_connectivity_2002,xuan_computing_2003}. Given two nodes $u$ and $v$ in a stream graph $S = (T,V,W,E)$, the latency from $u$ to $v$, denoted by $\ell(u,v)$, is the minimal time needed to reach $v$ from $u$ by following links of $S$ in a time-respecting manner, and taking into account node dynamics, see \cite{latapy_stream_2018} for details.

Notice that latencies in $S_\Delta$ are necessarily larger than or equal to latencies in $S$, since paths in $S_\Delta$ are also paths in $S$. Therefore, latencies in $S_\Delta$ are upper bounds of latencies in $S$, and we show below that they are actually quite close.

Figure~\ref{fig:mawi_lrmse_according_to_delta} displays the average difference between latencies in $S$ and $S_\Delta$ as a function of $\Delta$ for the Mawi dataset: $\frac{\sum_{u,v \in V, u\neq v}\ell_\Delta(u,v) - \ell(u,v)}{n\cdot(n-1)}$. It also displays the average latency stretch $\frac{\sum_{u,v \in V, u\neq v}(\ell_\Delta(u,v)+1) / (\ell(u,v)+1)}{n\cdot(n-1)}$ and the latency root mean square error:

$$ \LRMSE(S, S_\Delta) = \sqrt{
\frac{\underset{u,v \in  V, u \ne v}{\sum} (\ell(u,v)-\ell_\Delta(u,v))^2}{n(n-1)}
}$$

The figure also indicates the number of node pairs that were reachable in $S$ but became unreachable in this approximation. It appears that latencies are not significantly impacted by approximation, thus confirming that $S_\Delta$, despite its reduced number of strongly connected components, captures key information available in $S$. More precisely, only $11$ temporal paths disappear for $\Delta = \delta / 10^3$ and $115$ disappear for $\Delta = \delta / 10^2$, among a total number of $2,888,917$. The over-estimate of latencies is very small, with a $\LRMSE$ of $0.51$ and $1.61$, respectively. This has important consequences. For instance, one may compute latencies in $S_\Delta$ from its strongly connected components, which are much easier to compute and store than the ones of $S$, and obtain this way fast and accurate upper bounds (or approximations) of latencies in $S$, like we did here for the \textit{Mawilab} dataset.

\section{Related Work}
\label{sec-related-work}



We focus here on connected components defined in~\cite{latapy_stream_2018}, but other notions of connected components in dynamic graphs have been proposed.
Several rely on the notion of reachability, which, in most cases, induces components that may overlap and are NP-hard to enumerate, see for instance~\cite{Bhadra2003Complexity,gomez_calzado_connectivity_2015,huyghues_despointes_forte_2016,nicosia_components_2012}. This makes them quite different from the connected components considered here.


Akradi and Spirakis~\cite{akrida_verifying_2019} study and propose an algorithm for testing whether a given dynamic graph is connected at all times during a given time interval. If it is not connected, their algorithm looks for large connected components that exist for a long duration.
Vernet {\em et al.}~\cite{vernet_study_2020} propose an algorithm for computing all sets of nodes that remain connected for a given duration, and that are not dominated by other such sets.
Unlike our work, these papers do not partition the set of temporal nodes.




Finally, observing the size of largest components is classical, and Nicosia {\em et al.}~\cite{nicosia_components_2012} study them in time-varying graphs, with a connectivity based on reachability through temporal paths. They have a component for each node, which may overlap.


\section{Conclusion}
\label{sec:conclusion}

We proposed, implemented, and experimentally assessed a family of polynomial algorithms to compute the connected components of stream graphs. These algorithms handle streams of dozens of millions of events, and output connected components in a streaming fashion. This brings valuable information in practice, as we illustrate on a large-scale real-world dataset. We also propose a dataset approximation method making computations much faster while preserving key properties of the original data. Up to our knowledge, it is the first time that a partition of temporal nodes into connected components is computed at such scales.



A promising perspective is to enumerate connected components without listing them: one may for instance output only component size and duration in this way. Fully dynamic algorithms are particularly appealing to this regard, as their complexity is dominated by the explicit component listing.



\section*{Acknowledgements}
This work is funded in part by ANR (French National Agency of Research) through Limass project (grant ANR-19-CE23-0010) and FiT LabCom.

\bibliographystyle{splncs03}
{\footnotesize
\bibliography{references}
}

\end{document}